\documentclass[a4paper,english,english,english,english,english,preprint,aps]{revtex4}
\usepackage{pslatex}
\usepackage[T1]{fontenc}
\usepackage[latin1]{inputenc}
\usepackage{array}
\usepackage{subfigure}
\usepackage{graphicx}

\makeatletter

\providecommand{\tabularnewline}{\\}

\usepackage{pslatex}
\usepackage[T1]{fontenc}
\usepackage[latin1]{inputenc}
\usepackage{array}
\usepackage{subfigure}
\usepackage{graphicx}

\makeatletter

\usepackage[T1]{fontenc}
\usepackage[latin1]{inputenc}
\usepackage{array}
\usepackage{subfigure}
\usepackage{graphicx}

\makeatletter

\usepackage{pslatex}
\usepackage[T1]{fontenc}
\usepackage[latin1]{inputenc}
\usepackage{array}
\usepackage{subfigure}
\usepackage{graphicx}

\makeatletter

\usepackage[T1]{fontenc}
\usepackage[latin1]{inputenc}
\usepackage{array}
\usepackage{subfigure}
\usepackage{graphicx}

\makeatletter

\usepackage{babel}
\makeatother

\usepackage{babel}
\makeatother

\usepackage{babel}
\makeatother

\usepackage{babel}
\makeatother
\begin{document}

\title{Virtual-move Parallel Tempering}

\author{Ivan Coluzza}

\author{Daan Frenkel}

\affiliation{FOM Institute for Atomic and Molecular Physics, Kruislaan 407 1098
SJ Amsterdam , The Netherlands}

\begin{abstract}
We report a novel Monte Carlo scheme that greatly enhances the power
of parallel-tempering simulations. In this method, we boost the accumulation
of statistical averages by including information about all potential
parallel tempering trial moves, rather than just those trial moves
that are accepted. As a test, we compute the free-energy landscape
for conformational changes in simple model proteins. With the new
technique, the sampled region of the configurational space in which
the free-energy landscape could be reliably estimated, increases by
a factor 20. 
\end{abstract}
\maketitle
The exponential increase in the speed of computers during the past
decades has made it possible to perform simulations that were utterly
unfeasible one generation ago. But in many cases, the development
of more efficient algorithms has been at least as important.

One of the most widely used schemes to simulate many-body systems
is the Markov-chain Monte Carlo method (MCMC) that was introduced
in 1953 by Metropolis et al. \cite{Metropolis}. In this algorithm
the average properties of a system are estimated by performing a random
walk in the configurational space, where each state is sampled with
a frequency proportional to its Boltzmann weight. In the Metropolis
algorithm, this is achieved by attempting random moves from the current
state of the system to a new state. Depending on the ratio of the
Boltzmann weights of the new and the old states, these trial moves
may be either accepted or rejected. Metropolis et al. showed that
the acceptance probability of trial moves can be chosen such that
Boltzmann sampling is achieved.

One important application of the MC method is the estimation of the
Landau free energy $F$ of the system as function of some order parameter
\[
F\left(Q\right)=-kT\left[\ln P\left(Q\right)\right].\]
 There are many situations where the MCMC method does not yield an
accurate estimate of $F$, because it fails to explore configuration
space efficiently. This is, for instance, the case in {}``glassy''
systems that tend to get trapped for long times in small pockets of
configuration space. In the early 1990's the so-called parallel-tempering
(PT) technique was introduced to speed up the sampling in such systems~\cite{Parallel Tempering,Parallel Tempering 2,Parallel Tempering 3}.

In a parallel-tempering Monte Carlo simulation, $n$ simulations of
a particular model system are carried out in parallel at different
temperatures (or other at different values of some other thermodynamic
field, such as the chemical potential or a biasing potential). Each
of these copies of the system is called replica. In addition to the
regular MC trial moves, one occasionally attempts to swap the temperatures
of a pair of these systems (say $i$ and $j$). The swapping move
between temperature $i$ and $j$, is accepted or rejected according
to a criterion that guarantees detailed balance, e.g.: \begin{equation}
P_{acc}(ij)=\frac{e^{\Delta\beta_{ij}\Delta E_{ij}}}{1+e^{\Delta\beta_{ij}\Delta E_{ij}}}\label{paralletempering}\end{equation}
 where $\Delta\beta_{ij}$ is the difference of the inverse of swapping
temperatures, and $\Delta E_{ij}$ is the energy difference of the
two configurations. Although there are other valid acceptance rules,
we used the one in Eq.\ref{paralletempering} because it was easy
to implement.

To facilitate the sampling of high free-energy states,''difficult''
regions, we use the Adaptive Umbrella Sampling \cite{umbrella,umbrella 2}.
In this (iterative) scheme, a biasing potential is constructed using
the histogram of the states, sampled during an iteration as follows
\begin{equation}
W_{I}\left(Q,T\right)=W_{I-1}\left(Q,T\right)-a\ln\left(P_{I}\left(Q\right)\right),\label{eq:muticanonical}\end{equation}
 where $W$ is the biasing potential function of an order parameter
$Q$, $I$ is the iteration number, $a$ is a constant that controls
the rate of convergence of $W$ (a typical value for $a$ is $0.05$),
and $T$ is the temperature. After iteration, $W$ converges to the
Landau free energy. As a consequence, $P(Q)\sim\exp(-\beta F(q))\exp(W(Q))$
becomes essentially flat and the biased sampling explores a larger
fraction of the configuration space. During the MC sampling, we include
the bias, and only at the end of the simulation we compute the free
energy $F\left(Q\right)$ from \[
F\left(Q\right)=-kT\left[\ln P\left(Q\right)+W\left(Q,T\right)\right],\]
 where $P\left(Q\right)$ is the probability of observing a state
characterized by the order parameter $Q$, and $W\left(Q,T\right)$
is the biasing potential of the last iteration computed at temperature
$T$. Combined with Parallel Tempering, the acceptance rule for the
temperature swapping move is then \begin{eqnarray}
acc_{ij} & = & \frac{e^{\Delta\beta_{ij}\Delta E_{ij}+\Delta W_{ij}}}{1+e^{\Delta\beta_{ij}\Delta E_{ij}+\Delta W_{ij}}}\label{PT acceptance rule}\\
\Delta W_{ij} & = & W_{I}(Q_{i},T_{j})-W_{I}(Q_{j},T_{j})+\nonumber \\
 &  & W_{I}(Q_{j},T_{i})-W_{I}(Q_{i},T_{i})\label{eq:PT bias}\end{eqnarray}
 where $i$ and $j$ are replica indices, and $I$ is the iteration
number. We refer to this scheme as APT (Adaptive Parallel Tempering)
\cite{The Paper,Multicanonical Parallel Tempering}.

In the conventional MCMC method all information about rejected trial
moves is discarded. Recently one of us has proposed a scheme that
makes it possible to include the contributions of rejected configurations
in the sampling of averages~\cite{WasteR}. In the present paper,
we show how this approach can be used to increase the power of the
parallel-tempering scheme.

In this scheme, we only retain information about PT moves that have
been accepted. However, in the spirit of refs.~\cite{WasteR}, we
can include the contribution of all PT trial moves, irrespective of
whether they are accepted. The weight of the contribution of such
a virtual move is directly related to its acceptance probability.
For instance, if we use the symmetric acceptance rule for MC trial
moves, then the weights of the original and new (trial) state in the
sampling of virtual moves are given by \begin{eqnarray*}
P_{N} & = & \frac{e^{\Delta\beta\Delta E_{O\rightarrow N}+\Delta W_{O\rightarrow N}}}{1+e^{\Delta\beta\Delta E_{O\rightarrow N}+\Delta W_{O\rightarrow N}}}\\
P_{O} & = & \frac{1}{1+e^{\Delta\beta\Delta E_{O\rightarrow N}+\Delta W_{O\rightarrow N}}}\,,\end{eqnarray*}
 where $\Delta W_{O\rightarrow N}$ is defined in Eq.\ref{eq:PT bias}.
We are not limited to a single trial swap of state $i$ with a given
state $j$. Rather, we can include all possible trial swaps between
the temperature state $i$ and all $N-1$ remaining temperatures.
Our estimate for the contribution to the probability distribution
$P_{i}$ corresponding to temperature $i$ is then given by the following
sum \begin{eqnarray*}
P_{i}\left(Q\right) & = & \sum_{j=1}^{N-1}\left(\frac{1}{1+e^{\Delta\beta_{ij}\Delta E_{ij}+\Delta W_{ij}}}\right)\delta\left(Q_{i}-Q\right)+\\
 &  & \sum_{j=1}^{N-1}\left(\frac{e^{\Delta\beta_{ij}\Delta E_{ij}+\Delta W_{ij}}}{1+e^{\Delta\beta_{ij}\Delta E_{ij}+\Delta W_{ij}}}\right)\delta\left(Q_{j}-Q\right),\end{eqnarray*}
 where the delta functions select the configurations with order parameter
$Q$. As we now combine the Parallel tempering algorithm with a set
of parallel virtual moves, we refer to the present scheme as Virtual-move
Parallel Tempering (VMPT).

To measure the efficiency of VMPT, we computed the free energy landscape
of a simple lattice-protein model. In this model, interaction with
a substrate can induce a conformational change in the proteins. For
the same system we had already explored the use of the conventional
adaptive PT scheme~\cite{The Paper}.

\begin{figure}[!p]
\begin{center}\subfigure[]{\includegraphics[%
  width=0.30\paperwidth]{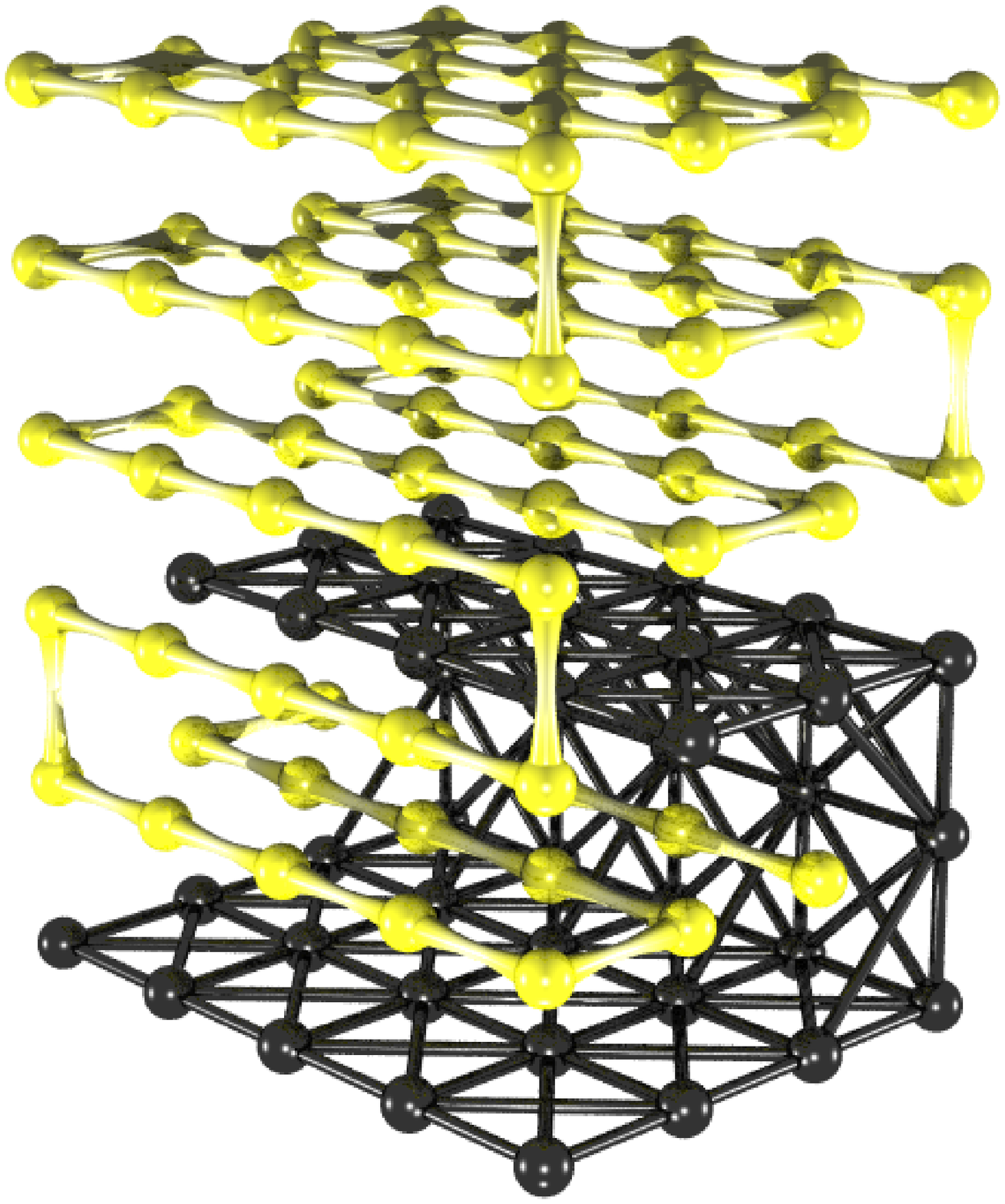}} \subfigure[]{\includegraphics[%
  width=0.30\paperwidth]{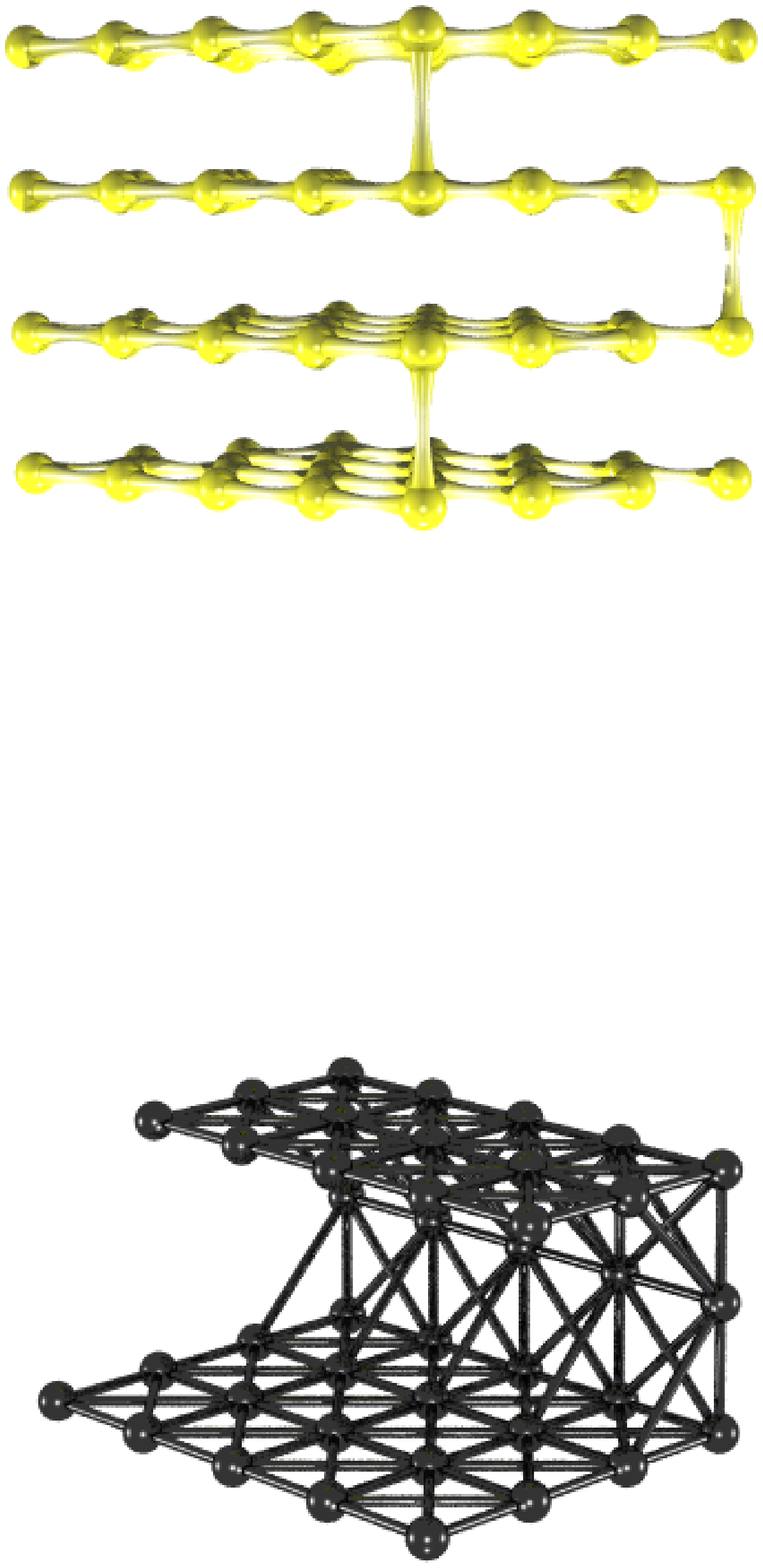}} \subfigure[]{\includegraphics[%
  scale=0.3]{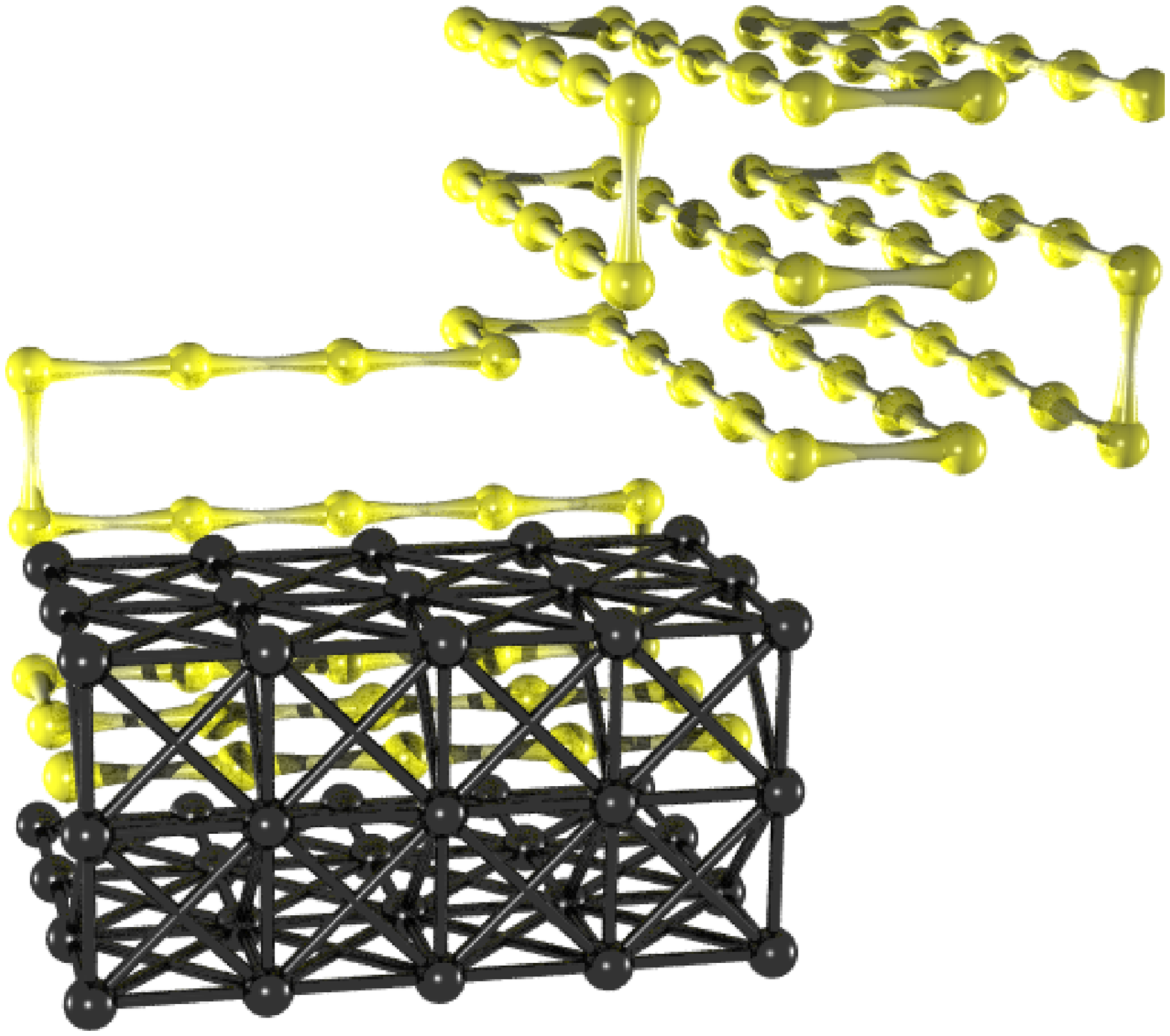}}\end{center}

\caption{(Color online) Spatial arrangement of the chain in the structures
used to test the model (a , b) , and intermediate structure ($Q=25$).\label{cap:Test-conformations}}
\end{figure}

Specifically, the model protein that we consider represents a heteropolymer
containing 80 amino acids, while the substrate has a fixed space arrangement
and contains 40 residues, see Fig.\ref{cap:Test-conformations}. The
configurational energy of the system is defined as \begin{equation}
E_{C}=E_{\textrm{intra}}+E_{\textrm{inter}}=\sum_{i}^{N_{C}}\left[\sum_{j\neq i}^{N_{C}}C_{ij}S_{ij}+\sum_{j'\neq i}^{N_{S}}C_{ij'}S_{ij'}\right],\label{Conformationalenergy}\end{equation}
 where the indices $i$ and $j$ run over the residues of the protein,
while $j'$ runs only over the elements of the substrate, $C$ is
the contact defined as \begin{equation}
C_{ij}=\left\{ \begin{array}{cc}
1 & \textrm{if i neighbor of j}\\
0 & \textrm{otherwise}\,.\end{array}\right.\end{equation}
 and $S_{ij}$ is the interaction matrix. For $S$ we use the 20 by
20 matrix fitted by Miyazawa and Jernigan \cite{S.Miyazawa} on the
basis of the frequency of contacts between each pair of amino acids
in nature.

We change the identity of the amino acids along the chain by {}``point
mutations'' which, in this context, means: changes of a single amino
acid. In doing so we explore the sequence space of the protein and
the substrate, and we minimize at the same time the configurational
energy of the system in two distinct configurations, one bound (Fig.\ref{cap:Test-conformations}.a)
and one unbound (Fig.\ref{cap:Test-conformations}.b). The design
scheme is the same as used in Ref.\cite{The Paper}. In this scheme,
trial mutations are accepted if the Monte Carlo acceptance criterion
is satisfied for both configurations.

The result of the design process is a model protein that has the ability
to change its conformation when bound to the substrate. The sampling
of the configurations is performed with three basic moves: corner-flip,
crankshaft, branch rotation. The corner-flip involves a rotation of
180 degrees of a given particle around the line joining its neighbors
along the chain. The crankshaft move is a rotation by 90 degrees of
two consecutive particles. A branch rotation is a turn, around a randomly
chosen pivot particle, of the whole section starting from the pivot
particle and going to the end of the chain. For all these moves we
use a symmetric acceptance rule with the addition of the biasing potential
calculated with the umbrella sampling scheme (Eq.\ref{eq:muticanonical})
\begin{equation}
acc_{O\rightarrow N}=\frac{e^{\beta\Delta E_{O\rightarrow N}+\Delta W_{O\rightarrow N}}}{1+e^{\beta\Delta E_{O\rightarrow N}+\Delta W_{O\rightarrow N}}},\label{eq:Moves acceptance}\end{equation}
 where $\Delta E_{O\rightarrow N}$ is the energy difference between
the new and the old state (Eq.\ref{Conformationalenergy}), and $\Delta W_{O\rightarrow N}$
is the difference in the bias potential from the same states (Eq.\ref{eq:muticanonical}).
We sample the free energy, as a function of two order parameters,
of which the first is the conformational energy defined in Eq.\ref{Conformationalenergy},
and the second is the difference of the number of contacts belonging
to two reference structures (e.g. 1 and 2) i.e. \begin{equation}
Q(C)=\sum_{i<j}^{N}\left[C_{ij}^{\left(1\right)}C_{ij}-C_{ij}^{\left(2\right)}C_{ij}\right]\,,\label{OrderParameter}\end{equation}
 where $C_{ij}^{\left(1\right)}$ and $C_{ij}^{\left(2\right)}$ are
the contact maps of the reference structures, and $C_{ij}$ is the
contact map of the instantaneous configuration. The order parameter
that measures the change in the number of native contacts is defined
as follows: as we consider two distinct native states, we take these
as the reference structures. Every contact that occurs to state 1
has a value and every contact that belongs to structure 2 has a value
$-1$. Contacts that appear in both $1$ and $2$, or do not appear
in neither of the two, do not contribute to the order parameter.

The reason why we assign negative values to native contacts of structure
2, is that we compute the free energy difference between the protein
in configuration 1 and 2. If we would have assigned 0 to the contacts
of structure 2 then we would not have been able to distinguish it
from unfolded configurations that do no have any native contacts at
all. For our specific case, $C_{ij}^{\left(1\right)}$ represents
the structure in Fig.\ref{cap:Test-conformations}.a, while $C_{ij}^{\left(2\right)}$corresponds
to the one shown in Fig.\ref{cap:Test-conformations}.b, and $Q$
has values between -15 and 30. Because the number of native contacts
includes the contacts with the substrate of the reference state, it
can be used to compute the free energy difference between the unbound
state and the specifically bound one.

\begin{figure}[!p]
\begin{center}\includegraphics[%
  width=0.50\paperwidth]{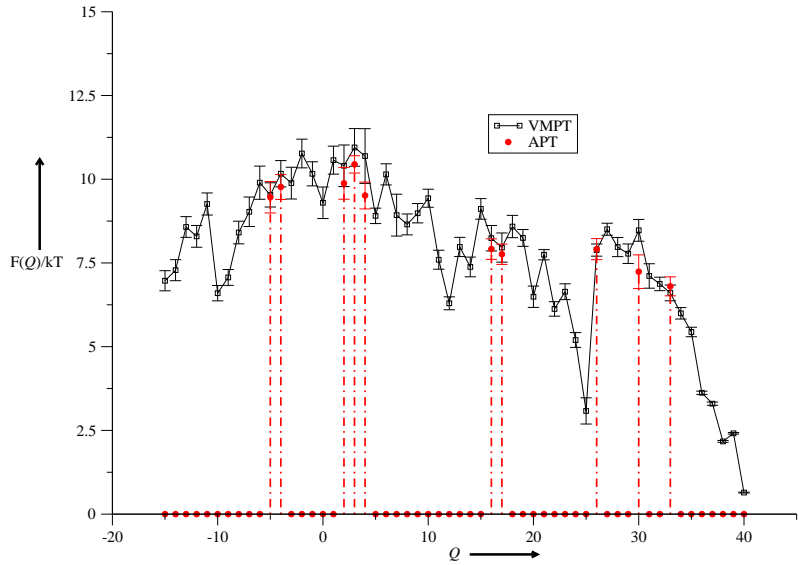}\end{center}

\caption{(Color online) Average free energy computed with 5 run (10\textasciicircum{}8
MC steps Tab.\ref{cap:ResumeTable}.I) of the old scheme, compared
with the result of 5 VMPT simulation ( 10\textasciicircum{}8 MC steps
Tab.\ref{cap:ResumeTable}.I), at $T=0.1$. The points with $F=0$
correspond to values of $Q$ that have not been sampled.\label{cap:Comparisong-1}}
\end{figure}

\begin{figure}[!p]
\begin{center}\subfigure[]{\includegraphics[%
  width=0.40\paperwidth,
  angle=-90]{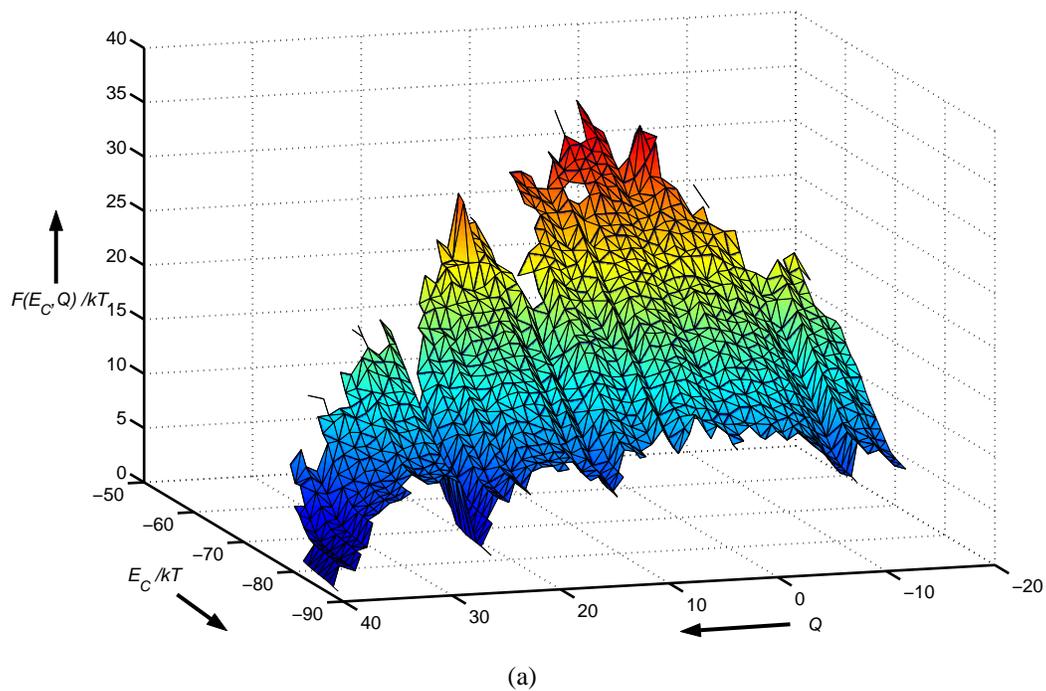}} \subfigure[]{\includegraphics[%
  width=0.40\paperwidth,
  angle=-90]{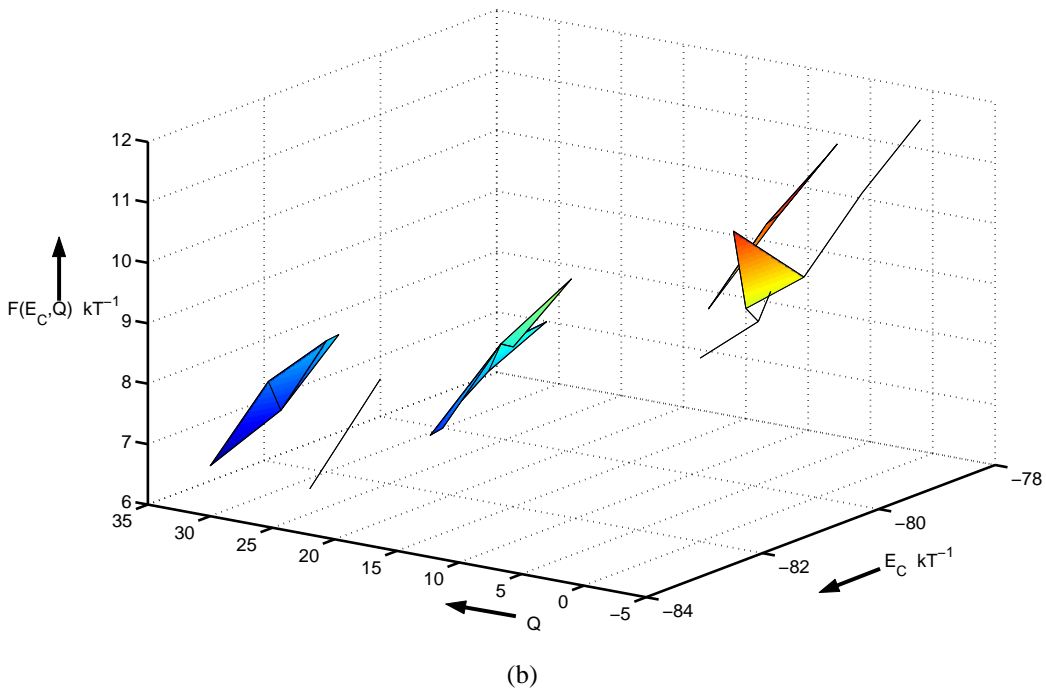}}\end{center}

\caption{(Color online) Plot of the free energy landscapes computed with the
algorithm VMPT (a) and the standard scheme APT (b). The free energies
$F(E_{C},Q)$ are function of the conformational energy $E_{C}$ (Eq.\ref{Conformationalenergy})
and of the number of native contacts $Q$ (Eq.\ref{OrderParameter}).
It is important to notice the big difference in the sampling, in fact
the number of points sampled with VMPT is 30 times bigger than the
one obtained with APT. \label{cap:3D-Comparison}}
\end{figure}

We performed 15 simulations, 5 of them with VMPT (using the parameters
in Tab.\ref{cap:ResumeTable}.I) and the other 10 with APT ( 5 using
the parameters in Tab.\ref{cap:ResumeTable}.I, and 5 with the parameters
in Tab.\ref{cap:ResumeTable}.II). In Fig.\ref{cap:Comparisong-1}
we compare the average free energies at $T=0.1$ (with error bars).
In this figure, we only show those free energies that were sampled
in all the 5 simulations of each group. From the figure it is clear
that the VMPT approach leads to a much better sampling of the free-energy
landscape. The advantage of the VMPT approach becomes even more obvious
if we plot the free energy {}``landscape'' as function of two order
parameters (viz. the conformational energy (Eq.\ref{Conformationalenergy})
and the number of native contacts). In this case the APT method is
almost useless as only small fragments of the free-energy landscape
can be reconstructed. The total number of points sampled with VMPT
is 20 times larger than with APT, and the energy range that is probed,
is one order of magnitude larger (see Fig.\ref{cap:3D-Comparison}).

\begin{figure}[!p]
\begin{center}\includegraphics[%
  width=0.50\paperwidth]{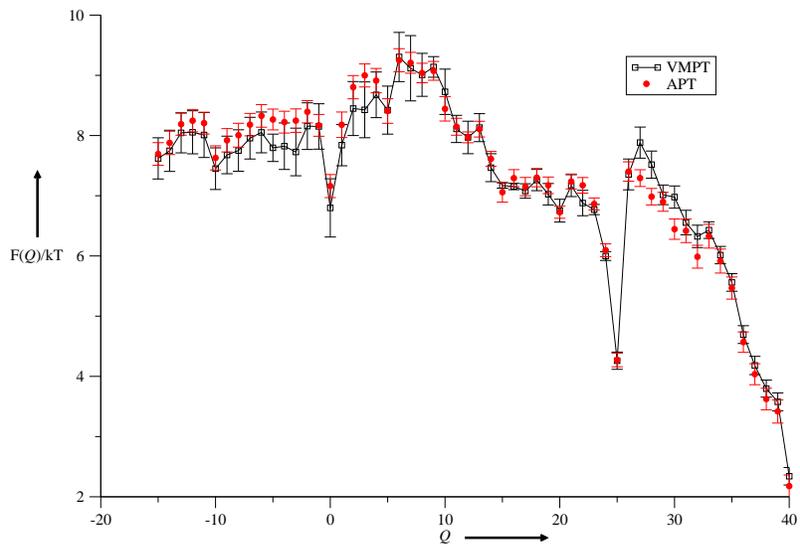}\end{center}

\caption{(Color online) Average free energy computed with 5 long run (10\textasciicircum{}10
MC steps Tab.\ref{cap:ResumeTable}.II) of the old scheme, compared
with the result of 5 shorter VMPT simulation ( 10\textasciicircum{}8
MC steps Tab.\ref{cap:ResumeTable}.I), at $T=0.5$\label{cap:Comparison-2}}
\end{figure}

To check the accuracy of the VMPT method, we compared the average
free energy obtained by APT and VMPT at high temperatures where the
APT scheme works reasonably well. As can be seen in Fig.\ref{cap:Comparison-2}
the two methods agree well in this regime (be it that a much longer
APT simulation was needed). Even though the APT runs required 20 times
more MC cycles, it still probes about 30\% less of the free-energy
landscape than the VMPT scheme.

As the implementation described above is not based on a particular
feature of the system under study, the results obtained in this study
suggest that the VMPT method may be useful for the study of any system
that is normally simulated using Parallel Tempering. Examples of the
application of Parallel Tempering in fully atomistic simulations of
protein folding can be found in refs.~\cite{Applica1,Applica2}.

\section*{Acknowledgments}

I. Coluzza would like to thank Dr. Georgios Boulougouris for many
enlightening discussions. This work is part of the research program
of the \char`\"{}Stichting voor Fundamenteel Onderzoek der Materie
(FOM)\char`\"{}, which is financially supported by the \char`\"{}Nederlandse
organisatie voor Wetenschappelijk Onderzoek (NWO)\char`\"{}. An
NCF grant of computer time on the TERAS supercomputer is gratefully
acknowledged.

\begin{table*}[!p]
\begin{tabular}{|c|p{5cm}|>{\centering}p{2cm}|>{\centering}p{2cm}|>{\centering}p{2cm}|>{\centering}p{2cm}|}
\hline 
Simulation&
\multicolumn{1}{c|}{Temperatures}&
Number of Iterations&
Sampling Steps&
APT exec time (sec)&
VMPT exec Time (sec)\tabularnewline
\hline
\hline 
I&
0.1 0.125 0.143 0.167 0.2 0.222 0.23 0.25 0.270000 0.29 0.31 0.33
0.350000 0.37 0.4 0.444 0.5 &
400&
$4\,10^{8}$&
2600&
3200\tabularnewline
\hline 
II&
0.1 0.125 0.143 0.167 0.2 0.222 0.23 0.25 0.270000 0.29 0.31 0.33
0.350000 0.37 0.4 0.444 0.5 &
1000&
$2\,10^{10}$&
150000&
\tabularnewline
\hline
\end{tabular}

\caption{Simulation parameters used for comparing the VMPT algorithm with
the old scheme. In Simul. I we used the same parameters for both algorithms.
The results in Fig.\ref{cap:Comparisong-1} show that VMPT was much
more efficient in sampling the free energy . In Simul .II, we increased
by two orders of magnitude the number of steps of the simulation with
APT to obtain a sampling of $F\left(Q\right)$comparable to the one
computed using the new VMPT scheme (Fig.\ref{cap:Comparison-2}).
Execution times computed on a SGI Altix 3700 with Intel Itanium II,
1,3 GHz \label{cap:ResumeTable}}
\end{table*}

\end{document}